\documentclass{jpsj-suppl}
\usepackage{txfonts} 
\usepackage{bm}

\newcommand{\be}{\begin{eqnarray}}
\newcommand{\ee}{\end{eqnarray}}

\newcommand{\ket}{\rangle}
\newcommand{\bra}{\langle}
\newcommand{\del}{\partial}

\newcommand{\Dslash}{{D\hspace{-7pt}/}}

\title{Charmed baryons and their interactions}

\author{Atsushi \textsc{Hosaka}$^{1,2,3}$,  
Emiko \textsc{Hiyama}$^{4,5}$, 
SangHo \textsc{Kim}$^{1,6}$, 
Hyun-Chul \textsc{Kim}$^{1,7}$, 
Hideko \textsc{Nagahiro}$^{8,1}$, 
Hiroyuki \textsc{Noumi}$^{1}$, 
Makoto \textsc{Oka}$^{5,3}$, 
Kotaro \textsc{Shirotori}$^{1}$, 
Tetsuya \textsc{Yoshida}$^{5}$, 
Shigehiro \textsc{Yasui}$^{5}$, 
}

\inst{
$^{1}$ Research Center for Nuclear Physics (RCNP), Osaka University, Ibaraki, 567-0031, Japan \\
$^{2} $J-PARC Branch, KEK Theory Center, Institute of Particle and Nuclear Studies,
KEK, Tokai, Ibaraki, 319-1106, Japan\\
$^{3}$ Advanced Science Research Center, Japan Atomic Energy Agency, Tokai, Ibaraki, 319-1195 Japan\\
$^{4}$ Nishina Center for Accelerator-Based Science, RIKEN, Wako 351-0198, Japan\\
$^{5}$ Department of Physics, Tokyo Institute of Technology, Meguro 152-8551, Japan\\
$^{6}$ APCTP, San 31 Hysa-dong, Namgu, Pohang Gyeongbuk 790-784, South Korea\\
$^{7}$ Department of Physics, Inha University, Incheon 402-751, Republic of Korea\\
$^{8}$ Department of Physics, Nara Womenfs University, Nara 630-8506, Japan\\
}

\email{hosaka@rcnp.osaka-u.ac.jp}

\abst{In this proceedings report, we discuss unique features of charmed, or in general heavy, baryons 
with one heavy quark.   
A well and long-term known phenomena, the distinction of the two modes of the $\rho$ and $\lambda$ type 
of a three-quark system is revisited.  
The difference of these modes may be tested in the production and decay reactions of the baryons which may be tested in the future experiments at J-PARC.}

\kword{Charmed baryons, $\rho$ and $\lambda$ modes, productions, decays}

\begin{document}
\maketitle

\section{Introduction}

Recent developments of hadron physics have been much motivated by the observations of 
exotic hadrons~\cite{hadron2013}.  
Many of the observations near the open heavy threshold 
imply that they may form the so called hadronic molecules. 
For instance, the $Z_b(10610)^+$ and $Z_b(10650)^+$ resonances which are 
almost on the $BB^*$ and $B^*B^*$ threshold are considered 
to be their molecular states~\cite{Belle:2011aa}.  
We may then imagine a situation where the four quarks $b, \bar b, u, \bar d$, 
which are the minimal contents of $Z_b^+$ mesons, are arranged to form 
a colorless mesonic clusters,  $B$ and $B^*$-mesons, forming a loosely bound state
by a relatively weak residual interaction between them.  
This could be a possible situation of what are expected to occur for four (in general multi) quark systems.  

Exotic mesons beyond the minimal configurations ($\bar qq$ for mesons and 
$qqq$ for baryons) have been already considered 
by Gell-Mann in his original paper of the quark model~\cite{GellMann:1964nj}. 
Multiquark systems may allow various types of correlations.
In $b, \bar b, u, \bar d$ of $Z_b^+$, colorless quark-antiquark correlation dominates.  
Less known is the colored diquark $qq$ correlation.  
Because they are confined just as quarks their properties are only indirectly 
approached.
Yet,  there have been many discussions for their evidence through  
such as neutron charge radius, $\Delta I$ = 1/2 rule in hyperon weak decays, 
$F_1/F_2$ ratio in the structure function, and so on~\cite{Lichtenberg:1980rv,Selem:2006nd}.  
These are phenomena in light quark systems (including the strange quark).  
Perhaps, another unique opportunity would be found in $qqQ$ heavy baryons, where
$Q$ is a heavy quark such as $c, b$.   
In the heavy quark limit, $Q$ behaves as a spectator with respect to 
the diquark $qq$, and so there will be chances to extract more clearly the properties of the $qq$ 
in the heavy baryons.  
They may be excited with respect to $Q$ or even their internal motions may be excited.    

Motivated by these backgrounds, a new experimental project has been discussed at J-PARC~\cite{e50}.
There the pion beam of high energy around 20 GeV will be available which can be used 
for the production of the charmed hadrons.  
In terms of their experimental developments of high resolution beam and spectrometer, 
they can measure 1 nb cross sections and even less in two-body (binary) reactions.  
These reactions enable us to study exclusively various excited states of charmed baryons.  
One advantage of such reactions is that the initial state of $\pi N$ is well controlled, and 
various reaction channels can be studied systematically and at once, and therefore,
for instance, relative properties of various matrix elements can be extracted.  

In this report we discuss  physics which can be studied in such reactions.  
Materials are based on the previous publications ~\cite{Kim:2014qha,Kim:2015ita} 
and currently progressing ones. 
In section 2, we revisit a distinctive features of the two 
different modes of the three-body system, $\lambda$ and $\rho$ type, 
which are degenerate in systems of equal quark masses (flavor SU(3) limit).  
In section 3, we discuss productions  of charmed baryons.
Production rates of various charmed baryons are discussed, where relatively 
large production rates for excited states are found.   
A similar feature in hyper nucleus production is pointed out.  
In section 4, we briefly discuss decays of charmed baryons, the work of which is currently going.  

\section{Mass spectrum}

\begin{figure}[h]
\begin{center}
\includegraphics[width=0.8 \linewidth]{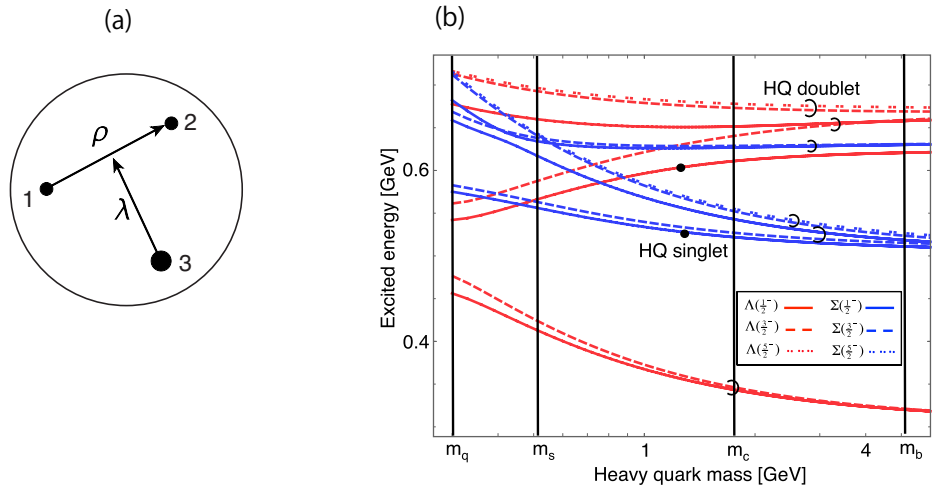}
\end{center}
\vspace{0mm}
\caption{(a) $\rho$ and $\lambda$ coordinates. (b) From Ref.~\cite{Yoshida:2015tia}; Heavy quark mass dependence of excited energy of first state, second state and third state for $1/2^{-}$(solid line), $3/2^{-}$(dashed line), $5/2^{-}$(double dotted line) of $\Lambda_Q$ (red line) and  $\Sigma_Q$ (blue line). Bullet denote heavy quark singlet. The pair within a semi-circle denote heavy quark doublet.}
\label{fig_masses}
\end{figure}

Let us consider baryons of three quarks which are confined by a harmonic oscillator 
potential and moving nonrelativistically.  
Let the mass of the two quarks $q$ denoted by $m$, and the one of the third quark $Q$ by $M$.  
The confinement force of harmonic oscillator is assumed to be independent of the flavor 
and is common for all three quarks.  
Such a Hamiltonian is  written as 
\be
H = \frac{\bm p_1^2}{2m} + \frac{\bm p_2}{2m} + \frac{\bm p_3}{2M}
+ \frac{k}{2} \left( (\bm x_1- \bm x_2)^2  + (\bm x_2- \bm x_3)^2  + (\bm x_3- \bm x_1)^2  \right) \ . 
\ee
By introducing the center of mass and Jacobi internal coordinates, 
$\bm \rho = \bm x_1 - \bm x_2$ and 
$\bm \lambda = ((\bm x_1 + \bm x_2)/2 - \bm x_3)$  (see Fig.~\ref{fig_masses} (a)), 
the above Hamiltonian can be separated into the parts of these variables.  
The frequencies for the $\lambda$ and $\rho$ modes are then given by 
\be
\omega_\lambda = \sqrt{\frac{M+2m}{M}} \omega\ , \; \; 
\omega_\rho = \sqrt{3} \omega\ ; \; \; \; \; 
\omega = \sqrt{\frac{k}{m}}\ .  
\ee
Therefore, when $M=m$ the two frequencies are the same,  
but when $M > m$, $\omega_\lambda < \omega_\rho$.
This is a consequence of the heavier inertia mass for the $\lambda$ motion 
than for the $\rho$ motion.
This feature is general for a potential which is milder 
than the Coulomb force ($\sim 1/r$) at the origin.  
Intuitively, excitations of heavier particle have less kinetic energy than that of a lighter particle.  
The separation of the two frequencies is a kinematic effect, 
which is know as the isotope shift~\cite{Copley:1979wj}.  
In the light quark systems of $uds$, typical excitation energies of the first excited states 
are of order $500 \sim 600$ MeV, while in the charmed baryons they are substantially smaller.
This implies that in the charmed baryons the two modes appear distinctively.  
Such mode separation may have already been appeared in 
the so-called $\Lambda$-$\Sigma$ inversion 
in the $J^P = 5/2^-$ states, as pointed out in Ref.~\cite{Copley:1979wj}
if the mass of the strange quark is sufficiently heavier than $u,d$ quark masses.  

Theoretically, it is interesting to see various baryon masses as functions 
of the heavy quark mass $M$ which can be smoothly varied.  
In Ref.~\cite{Yoshida:2015tia}, we have performed such calculations 
using  more realistic Hamiltonian with linear confinement force and 
other hyper fine (residual) interactions.  
The essential features as discussed above by using the harmonic oscillator potential, 
however, do not change.  

In Fig.~\ref{fig_masses}, 
excitation energies of various $\Lambda$ and $\Sigma$ baryons are shown as functions 
of the heavy quark mass $M$ which is varied from the SU(3) limit of light flavors ($m_u = m_d = M$) 
to the heavy quark limit ($M \to \infty$).  
In these calculations, various residual interactions resolve the degeneracy 
seen in the simple harmonic oscillator model both in the SU(3) and heavy quark limits.  
Following features are observed.
\begin{itemize}

\item 
In the entire $M$ region, the lower two states are well separated from the others.  
These are the $\Lambda$ particles of $\lambda$ modes.  

\item
In the SU(3) limit at $M = m_q$, 
various states are mixtures of the $\lambda$ and $\rho$ modes and seem 
to show some complicated structure.  
Accordingly, the wave functions are mixtures of the $\rho$ and $\lambda$ modes.  

\item 
As $M$ is increased and becomes larger than 1 GeV, $\lambda$ and $\rho$ modes started to separate; 
all seven $\lambda$ modes become lower than the other $\rho$ modes.  

\item 
In the heavy quark limit, 
in the presence of the residual interactions as in the case shown in Fig.~\ref{fig_masses}, 
the lower seven lambda modes split into two and five states.
This is due to the spin-spin force  
for the $\Lambda$ and $\Sigma$ states, separating the so called good and bad diquark masses.  
In the absence of the residual interaction, however, 
these seven states become degenerate.  

\end{itemize}

Generally it seems that the mass spectrum becomes simpler in the heavy quark limit than in the SU(3) limit.  
In particular, in the heavy quark limit, there appear the heavy quark doublet or singlet, which is a consequence of 
heavy quark spin symmetry.  
In the present case the total baryon spin $J$ is the sum of the spins of the (light) diquark $j_l$ and the heavy quark $j_Q$.  
The diquark spin takes either 0 or 1, which generate the singlet and doublet respectively.  
In Fig.~\ref{fig_masses}, heavy quark doublet is indicated by the semi-circle and the singlet by the blob.  
Lastly, 
the $\Lambda$-$\Sigma$ inversion is seen immediately above the SU(3) limit as the mass $M$ starts to increase from $m_q$, 
which is denoted by the double dotted lines.

\section{Productions in a Regge model}

Production rates, or cross section values, are relevant issues when discussing 
new experimental projects.  
At J-PARC, the charmed baryon physics has been discussed and is on their plan 
by using the high momentum pion beam of about 20 GeV.  
This energy is kinematically sufficient to produce charmed baryons together with 
a $D$ or $D^*$ meson, up to the excitation energies of about 1 GeV.  
Theoretically, however, it is difficult to make reliable estimations or predictions 
of the production rates in this energy region.  
So far there is one experimental data for the pion induced reactions
at BNL in mid 80's, which reported only the upper limit of $\sim$ 7 nb~\cite{Christenson:1985ms}.  
Under the present days improved conditions, it has been simulated and shown that 
cross sections of order 1 nb is possible to be measured to extract 
signals of charmed baryon productions.   
In fact, it would be possible even with a smaller rate~\cite{e50}.  
Thus it is desired to have a theoretical predictions with reasonable reliability.  

Experimentally, they plan to perform measurements of charmed baryon productions 
associated with a vector $D^*$ meson, rather than a pseudoscalar $D$ meson
to obtain better signals.  
Thus we are interested in the binary reaction 
\be
\pi + N \to D^* + Y_c\ , 
\label{eq_piNtoBY}
\ee
where $Y_c$ is a charmed baryon (either ground or excited state).  

An analogous reaction in the strangeness sector is $\pi + N \to K^* + Y$ with $Y$ a ground state strange hyperon
(either $\Lambda$ or $\Sigma$),
for which there are some experimental data~\cite{Dahl:1967pg,Crennell:1972km}.  
Angular distributions of this reaction at the energy region from the threshold to a few GeV above  
show a typical diffractive pattern of forward peak~\cite{Crennell:1972km}, 
indicating the dynamics dominated by the $t$-channel one.   
This is the situation where the Regge approach can be applied.  
In Ref.~\cite{Kim:2015ita}, 
we have examined carefully the cross sections of the strangeness production reactions~\cite{Kim:2015ita}.  

\begin{figure}[h]
\begin{center}
\includegraphics[width=0.9 \linewidth]{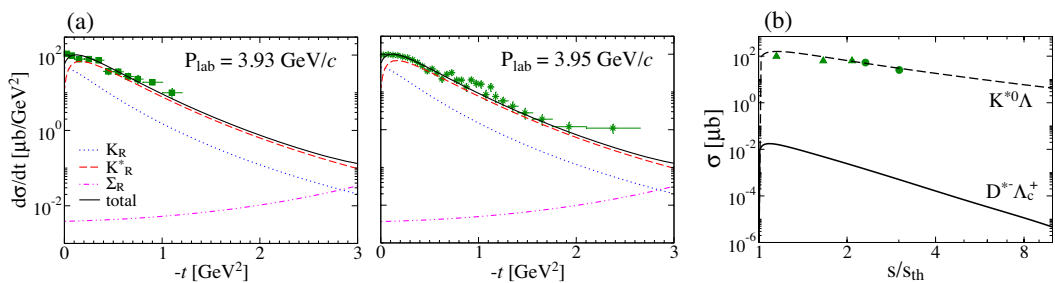}
\end{center}
\vspace{0mm}
\caption{(a) Differential cross sections for $\pi N \to K^* \Lambda$ as functions of $-t$ and 
(b) total cross sections for $\pi N \to K^* \Lambda$ and $\pi N \to D^* \Lambda_c$ as functions of the energy normalized by the threshold energy $s_{\rm th}$.  }
\label{fig_piNXsections}
\end{figure}

In that approach, one important but undetermined parameter theoretically is the normalization factor for 
absolute values, for which we must rely on experimental data at a certain point of energy and/or angle.  
Once it is determined, then, the resulting energy and angular dependence are predicted by the model.  
It has turned out that the dominant contribution is made by the vector ($K^*$) Reggeon exchange which 
reproduces both the angular ($t$)  and energy dependences.  
The results are shown in Fig.~\ref{fig_piNXsections}.  
Differential cross sections for  $\pi N \to K^* \Lambda$ in Fig.~\ref{fig_piNXsections} (a) 
are compared with experimental data
which is nicely reproduced by our approach after adding various contributions, 
$K^*$, $K$, and $\Sigma$ Regeon exchanges in the $t$, $t$ and $u$-channels, respectively.  
We emphasize that the cross section is dominated by the vector ($K^*$) Reggeon, 
which determines large part of the $-t$ dependence.  
At small $-t$ of forward scattering, however, the interference between the $K^*$ and $K$
Regeons is important.   
The total cross section as functions of the energy (momentum) 
is shown in Fig.~\ref{fig_piNXsections}~(b) which compares well with the data.  
Note that the slope of the total cross section is determined by the intercept of the Regge trajectory and is 
consistent with unitarity.
This is an advantage over the effective Lagrangian approach which is also used often in literatures~\cite{Kim:2015ita}.  

Having established a good description for the strangeness productions, 
we may be able to extend it to the charm productions, by assuming that the strange quark is 
replaced by the charm quark without changing the dynamics much.  
Note that this is indeed an assumption and therefore, the results coming out should be tested 
by experiments.  
The results are shown in Fig.~\ref{fig_piNXsections}~(b), where the total cross section is drawn 
in comparison with the strangeness production.  
As compared to the strangeness production the charm production rate is suppressed by about $10^4$ 
near the threshold and even further as energy is increased.  
As a consequence the total cross section at the energy of the BNL experiment, 
$p_{\rm lab} \sim 13$ GeV, is about 10 nb, which is the same order of their upper limit.  
In our calculations, there is an ambiguity in the choice of coupling strengths, form factors and so on, 
while much part of them are reasonably determined.  
There remains still an ambiguity in the absolute values within several factors.  
It is a challenge to reduce the ambiguity in future theoretical studies.  

\section{Productions of various baryons}

Once we have estimated the cross section of the ground state $\Lambda_c$, we would like to study 
production rates of various charmed baryons, not only the ground state 
but also excited states~\cite{Kim:2014qha}.   
These production rates reflect the information of the structure, and is very important for the study 
of the charmed baryons.  
Theoretically,  we need to have the wave functions and to specify reaction mechanism.  
As we have discussed in the previous section, the charmed baryon production reactions are expected to be 
dominated by diffractive $t$-channel dynamics, in particular the vector Reggeon exchanges.  
Therefore, in the following, we estimate the forward cross sections where the helicity conservation 
brings a unique structure in spin matrix elements and make the actual computation simple also.  

\begin{figure}[h]
\begin{center}
\includegraphics[width=0.8 \linewidth]{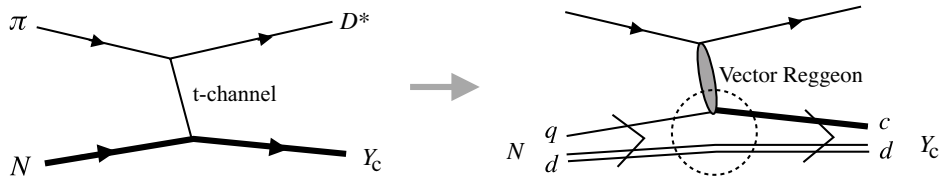}
\end{center}
\vspace{0mm}
\caption{Quark diagram for charmed baryon productions.}
\label{fig_piNDY}
\end{figure}

In a microscopic picture, as shown in Fig.~\ref{fig_piNDY}, 
the incoming pion exchanges the $D^*$ Reggeon which 
is absorbed by a light quark in the nucleon to the charm quark in a charmed baryon.  
In this process the remaining two quarks denoted by $d$ ($\sim$ diquark) are assumed to be 
intact behaving as a spectator.  
Thus this is essentially the one-body process.  
Assuming further that the vector  Reggeon is dominated by the $D^*$ vector meson, 
we can write the vertex structure for $\pi D^* D^*$ and $D^*qc$ 
($q$: light quark and $c$: charm quark)~\cite{Kim:2014qha}.  
For simplicity, we assume that the $D^*qc$ vertex is dominated by the Dirac term, 
the product of the vector current and the vector $D^*$ meson ($\bar c \Dslash^{\ *}q$) .  
An advantage of the present study is, as in the previous section, 
that we can apply the method
not only to the charm sector but also to other flavor sectors, 
assuming the similar mechanism.  
Here we will see the comparison between the charm and strange sectors.  

Because the $\pi D^*D^*$ vertex is common, 
the production rates $R(Y_c)$ to various charmed baryons $Y_c$
are given by the matrix elements of the $D^*NY_c$, which correspond in Fig.~\ref{fig_piNDY}
to the vertex circled by the dashed line,  
\be
R(Y_c) \sim |t_{fi}|^2\ , \; \; \; 
t_{fi} = \bra Y_c | {\cal O} | N \ket
\ee
where $|N\ket$ and $|Y_c\ket$ are the initial nucleon and  a final charmed baryon.
For $Y_c$, we consider not only the ground state but also various excited states of angular momentum 
up to $l=2$, thus including the $p$- and $d$-wave baryons.  
Detailed calculation is explained in Ref.~\cite{Kim:2014qha}.  
Here we discuss the essential part that it is proportional to 
\be
t_{fi} \sim \bra Y_c | \bm e_\perp \cdot \bm \sigma \exp(i \bm q_{\rm eff} \cdot \bm x) | N\ket \ , 
\label{t_fi}
\ee
where $\bm q_{\rm eff}$ is a momentum fraction transferred to the quark that is excited, 
which is a function of the quark masses and the number of the quarks in the baryon, 
and 
$\bm e_\perp$ is the transverse polarization of the outgoing $D^*$ meson.  
The presence of the transverse component is related to the conservation of the helicity 
in the forward scattering amplitude, and is characteristic for vector meson exchange process.  
If, for instance, pseudoscalar Reggeon is exchanged instead, 
it is replaced by the longitudinal spin, which is parallel to the momentum.  

One remark is that in the present production mechanism which is dominated by the one-body operator, 
only the $\lambda$-modes are excited.
For the $\rho$-mode excitations, we need more complicated process beyond the one-body mechanism.   
To go beyond the one-body process is an important problem to be considered in the future. 

The matrix elements $t_{fi}$ are decomposed into a geometrical factor which is  
expressed by the Clebsch-Gordan coefficients, and a dynamical factor which is given by the 
Fourier transform of  the transition density.  
The former geometric factor is useful to identify like-transitions such as the ones to heavy quark spin partners,
while the latter reflects the dynamical information of the baryon wave functions which is particularly 
sensitive to the angular momentum of the excited baryons. 

We have performed the computation of the matrix elements by using the harmonic oscillator wave functions.  
Explicit forms of wave functions for various baryons are given in Ref.~\cite{Kim:2014qha}.  
Numerical calculations are rather straightforward, 
and the results are summarized in Table~\ref{Transitions}, 
where shown are not only for the charmed baryons but also 
for the corresponding strange baryons.  
The comparison between them reveals characteristic feature of the production rates.  
To see them first we show the essential part for the analytic expression for dynamical part of the amplitudes
for $l=0, 1, 2$ as 
\be
\bra Y_c(s,p,d\ {\rm wave})| \bm e_\perp \cdot \bm \sigma e^{i \bm q_{eff} \cdot \bm x} | N(s\ {\rm wave}) \ket 
\sim \left( \frac{q_{eff}}{A} \right)^l \times \exp\left( - \frac{q_{eff}^2}{4A^2}\right)
\label{transitionME} 
\ee
where $l = 0, 1, 2$ for $s, p, d$ wave baryons respectively, 
$A$ is the average of the oscillator wave function parameters for the initial and final baryons
$
A^2 = ({\alpha^2 + \alpha^{\prime 2}})/{2}
$,
and 
$\alpha = \sqrt{m\omega}$.  
An interesting and important feature of this formula (\ref{transitionME}) is 
that depending on the values of $q_{eff}$, 
most productive rate is obtained for different $l$.  
$A$ which is the parameter for the baryon extension is not 
very dependent on the mass of  the single heavy quark (either charm or strange).  
On the other hand, $q_{eff}$ depends much on it; 
for the productions of strange baryons $q_{eff} \sim $ several hundred MeV, while 
for the productions of charmed baryons $q_{eff} \sim $ one GeV.  
Therefore, for strange baryons the ground state ($l=0$) is produced the most, 
while for charmed baryons, excited states of $l \sim 1$ are more produced. 
In physical term, better angular momentum matching is achieved for larger $l$ for  
larger $q_{eff}$ for charm productions.  
This implied that there is more chance to study excited states for charmed baryons.  
That larger $l$ states are produced more than lower $l$ states has been observed 
in hyper nucleus productions~\cite{Hotchi:2001rx}.
There the hyper nucleus has a larger size of order a few fm, and so this is realized for lower 
momentum transfer.  
Another aspect to be  pointed out is due to the geometric factor which relates 
the production rates of the heavy quark spin partners.  
For instance, the ratio of the $J^P=1/2^-, 3/2^-$ partner production is exactly 1/2.  

\begin{figure}[h]
\begin{center}
\includegraphics[width=0.5 \linewidth]{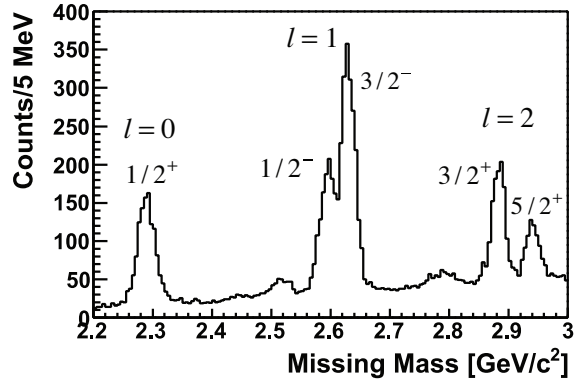}
\end{center}
\vspace{0mm}
\caption{Numerical simulations for the $\Lambda_c$ baryons of $l=0, 1, 2$ using the results given 
in Table~\ref{C_and _R}.   }
\label{fig_simulation}
\end{figure}

To illustrate how the excited states are observed in actual experiments, 
in Fig.~\ref{fig_simulation}, we show the results of numerical simulation 
for expected spectrum using the results of Table~\ref{C_and _R} and expected experimental 
observations with possible background subtracted.  
It is clearly seen that excited states of $l=1, 2$ are more (similarly) produced than (to) the ground state
of $l = 0$, 
as well as the ratios of the heavy quark doublet $1/2^-, 3/2^-$ and $3/2^+, 5/2^+$ given 
by the geometric factors.   

\begin{table}[htdp]
\caption{Production rates of various excited charmed/strange baryons normalized by the one of the ground state.
Masses of excited states are tentatively assigned if there are data, otherwise given a simple value.  
They are computed 
at $k_\pi^{Lab} = 4.2$ GeV for the strange, and 
at $k_\pi^{Lab} = 20$ GeV for the charmed baryons.}
\begin{center}
\begin{tabular}{c c c c c c c c c }
\hline
\hline
$l =0$ & 
$\Lambda(\frac{1}{2}^+)$ & 
$\Sigma(\frac{1}{2}^+)$ & 
$\Sigma(\frac{3}{2}^+)$ & 
 & & & &  \\
 $M$ [MeV]  & 1116 & 1192 & 1385 &  &  &  &  &   \\
    & 2286 & 2455 & 2520 &  &  &  &  &   \\
\hline
 %
${\cal R}(B_s)$  & 1 & 0.04 & 0.210 & & & & &   \\
${\cal R}(B_c)$  & 1 & 0.03 & 0.17 & & & & &   \\
 \hline
\hline
$l = 1$ & 
$\Lambda(\frac{1}{2}^-)$ & 
$\Lambda(\frac{3}{2}^-)$ & 
$\Sigma(\frac{1}{2}^-)$ & 
$\Sigma(\frac{3}{2}^-)$ & 
$\Sigma^\prime(\frac{1}{2}^-)$ & 
$\Sigma^\prime(\frac{3}{2}^-)$& 
$\Sigma^\prime(\frac{5}{2}^-)$  & \\
$M$ [MeV] & 1405 & 1520 & 1670 & 1690 & 1750 & 1750 & 1775 & \\
     & 2595 & 2625 & 2750 & 2800 & 2750 & 2820 & 2820 & \\
\hline
%
${\cal R}(B_s)$ & 0.07 & 0.11 & 0.002 & 0.003 & 0.003 & 0.01 & 0.01 & \\
${\cal R}(B_c)$ & 0.93 & 1.75 & 0.02 & 0.04 & 0.05 & 0.21 & 0.21 & \\
\hline
\hline
$l = 2$ & 
$\Lambda(\frac{3}{2}^+)$ & 
$\Lambda(\frac{5}{2}^+)$ & 
$\Sigma(\frac{3}{2}^+)$ & 
$\Sigma(\frac{5}{2}^+)$ & 
$\Sigma^\prime(\frac{1}{2}^+)$ & 
$\Sigma^\prime(\frac{3}{2}^+)$& 
$\Sigma^\prime(\frac{5}{2}^+)$ &
$\Sigma^\prime(\frac{7}{2}^+)$ \\
$M$ [MeV] & 1890 & 1820 & 1840 & 1915 & 1880 & 2000$^*$ & 2000$^*$ & 2000$^*$ \\
      & 2940 & 2880 & 1840 & 3000$^*$ & 3000$^*$ & 3000$^*$ & 3000$^*$ & 3000$^*$ \\
\hline
${\cal R}(B_s)$ & 0.02 & 0.04 & 0.003 & 0.001 & 0.001 & 0.001 & 0.001 & 0.001 \\
${\cal R}(B_c)$ & 0.49 & 0.86 & 0.01 & 0.02 & 0.01 & 0.05 & 0.11 & 0.09 \\
\hline
\end{tabular}
\end{center}
\label{C_and _R}
\end{table}%

\section{Decays of low lying excited states}

\begin{figure}[h]
\begin{center}
\includegraphics[width=0.9 \linewidth]{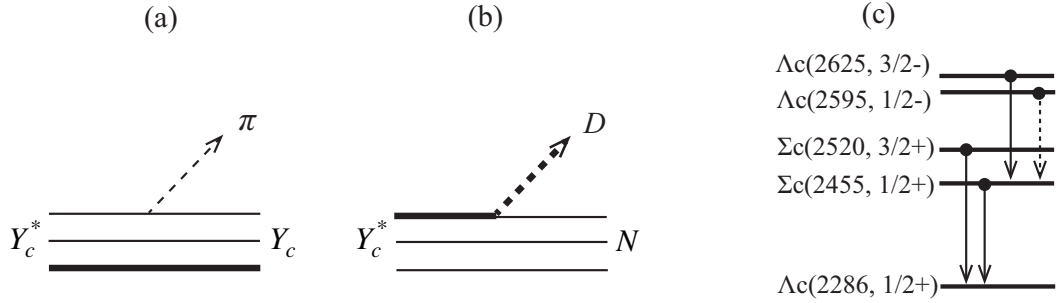}
\end{center}
\vspace{0mm}
\caption{Decays of charmed baryons.  (a) and (b) are decays into $\pi + Y_c$ (pion emission) and $D + N$, 
respectively.  (c) shows possible pion emissions between various states.  Dashed arrow indicates that the 
decay occurs only through the tale of a finite width of resonances.}
\label{fig_decays}
\end{figure}

We are currently working on the decays, and therefore, we only present 
some unique features of the decays of charmed baryons.  
Detailed results will be reported elsewhere.  
There are two two-body decay modes of $Y_c^*$; 
one is to $\pi + Y_c$ (pion emission, Fig.~\ref{fig_decays} (a)) and the other to $D + N$ (Fig.~\ref{fig_decays} (b)) .  
In the quark model, it is described  by one-body operator 
of the meson and quark interaction, either $\pi qq$ or $D qQ$, 
where $q$ and $Q$ denote light ($u,d$) and heavy ($c$) quarks, respectively.  
When the one-body process dominates, the $\lambda$ modes decay into 
both $\pi + Y_c$ and $D + N$, 
while the $\rho$ modes into only $\pi + Y_c$.  

For the pion emission, an interesting situation occurs, 
where the momentum (energy) of the emitted pion 
becomes very low, because of relatively small excitation energies of $Y_c^*$.
The relevant transitions are shown in Fig.~\ref{fig_decays} (c), where typically 
the pion momentum is of order about 100 MeV.
In particular, $\Lambda_c(2595)$ decays into $\Sigma_c(2455)$ with null 
pion momentum if central values of their masses are taken.  
$\Lambda_c(2595)$ can decay only between the tail part of the two finite widths.  
This situation contrasts the light flavor case, where in many decays the emitted pion 
carries several hundred MeV; for instance, even the lowest resonance $\Delta$ decay is 
associated wit the pion of momentum $p \sim 230$ MeV.  

Because of this these transitions may provide a good opportunity to test 
low energy behaviors of the pion interacting with the light component
of the heavy baryons, in particular with constituent quarks.  
In this regard, 
it is interesting to look at possible forms of the basic pion-quark interaction.  
Generally, there are two independent terms for the $\pi qq$ interaction, 
pseudoscalar and pseudovector types, 
\be
 i g \bar q \gamma_5 \bm \tau \cdot \bm \pi q\ , \; \; \; 
\frac{g_A}{f_\pi}  
\bar q \gamma_\mu \gamma_5 \bm \tau \cdot \del^\mu \bm \pi q\ .
\ee
These two interactions are equivalent each other for the matrix elements of the on-shell quarks 
for the initial and final states, but are generally different for off-shell quarks which is the case 
for those confined in a baryon.   
In non-relativistic calculations the two independent terms are given as 
\be
\bm \sigma \cdot \bm q \ , \; \; \; 
\bm \sigma \cdot \bm p_i \ ({\rm or}\ \bm \sigma \cdot \bm p_f)
\label{H_piqq_NR}
\ee
where $\bm p_{i}, \bm p_{f}$ are the initial and final momentum and 
$\bm q = \bm p_f - \bm p_i$.
The pseudoscalar  or $\bm \sigma \cdot \bm q$ type interaction is suppressed 
when  $\bm q$ becomes zero as in the case for the decay of 
$\Lambda_c(2595)$ into $\Sigma_c(2455)$, while 
the pseudovector or $\bm \sigma \cdot \bm p_i$ one can contribute, which indeed 
explains that decay.  
Moreover, spontaneous break down of chiral symmetry which implies non-linear realization 
naturally leads to the derivative coupling of pseudovector type.
The non-relativistic reduction of the pseudovector type interaction leads to 
both two terms of (\ref{H_piqq_NR}) with the strengths determined.  
Thus the use of the pseudovector type interaction is a good testing ground 
of chiral symmetry with spontaneous break down for the pion and constituent quarks.  
A preliminary results indicate that this method provides a good descriptions for the decays 
shown in Fig.~\ref{fig_decays} (c) which will be reported elsewhere.

\section{Summary}

In this proceedings paper, we have reported our recent activities 
for the charmed baryons, which will be tested in the future J-PARC experiments.  
The role of a heavy quark is to disentangle 
the two degrees of freedom of $\rho$ and $\lambda$ modes, 
providing clearcut understanding of these modes of quarks as building blocks.  
The disentanglement  is primarily due to the mass difference, kinematic isotope effect, 
and is a universal phenomena independent of the details of the dynamics.  
To verify these features experimentally, we have estimated the productions rate of 
various charmed baryons including excites states.  
We have also reported the progress status of our decay study.  

The results are summarized as follows

\begin{itemize}

\item
The $\lambda$ modes are lowered as compared to the $\rho$ modes, because of the collectivity 
of the diquark motion rather than a single quark motion.  

\item
The Regge model predicts the charm production rate in the pion induced reactions near the threshold 
around a few [nb] which is well tested in the future J-PARC experiments.  

\item
The production mechanism of one-body process dominates the formation of the $\lambda$ modes.
The resulting production rates of excited states are expected to be large due to angular momentum 
matching in large momentum transfer reactions.  

\item
In pion emission decays  the pion momentum is 
very low as compared to the decays of light flavor baryons.  
This can be a good testing ground for low energy chiral dynamics between quarks and pions.  

\end{itemize}

There are several important issues to be studied in the future.  
One is to establish a more reliable theoretical approach to the dynamics of charm quark production
in a quantitative manner.   
This is interesting by itself but is also important for the study of baryon structure.  
Another issue is the identification of various states predicted by the theory
with experimental data and clarify the nature of the excitations.  
Lastly, because comparisons are made with  scattering experiments, reaction mechanism 
should be clarified.  
In the present study, our considerations have been limited to one-body processes.  
It would be very important to go beyond multi-body processes which is needed to reveal 
actual correlations between quarks in multi-quark systems.  

Though we have not so far discussed explicitly, 
all these considerations must be also tested by QCD.  
There is a significant progress in hadron spectroscopy in lattice QCD.  
The charmed baryons are directly investigated for the ground states~\cite{Namekawa:2013vu}.
In the light flavor sector, excited states have been also studied~\cite{Edwards:2011jj,Roberts:2013oea}.  
The excited states couple with open hadron channels where the dynamics of the hadrons 
becomes important.  
This is related to the problem of the reaction dynamics in scattering experiments.  
In this regard, the development in the study of hadron-hadron interactions is  
also important~\cite{Ishii:2006ec}.  
Thus mutual approaches of theories and  experiments 
should  work coherently toward the understanding the physics of heavy baryons and of QCD.  

\section*{Acknowledgements}

We thank Sasha Titov and Katsunori Sadato for useful discussions.  
AH acknowledges the support in part by Grant-in-Aid for Science Research (C) 26400273  by the MEXT.  


\end{document}